\documentclass[11pt,a4paper]{article}
\usepackage{jheppub}
\usepackage{subfig}

\newcommand{\ba}{\begin{eqnarray}}
\newcommand{\ea}{\end{eqnarray}}

\newcommand{\be}{\begin{equation}}
\newcommand{\ee}{\end{equation}}
\newcommand{\bea}{\begin{eqnarray}}
\newcommand{\eea}{\end{eqnarray}}
\newcommand{\f}{\frac}

\newcommand{\nn}{\nonumber}

\title{On using Cold Baryogenesis to constrain the Two-Higgs Doublet Model.}

\author[a]{Anders Tranberg,}
\author[b]{Bin Wu}

\affiliation[a]{Niels Bohr International Academy and Discovery Center, Niels Bohr Institute,\\  Blegdamsvej 17, 2100 Copenhagen, Denmark}
\affiliation[b]{Faculty of Physics, University of Bielefeld, Bielefeld, Germany}

\emailAdd{anders.tranberg@nbi.dk}
\emailAdd{binwu@physik.uni-bielefeld.de}

\abstract{We consider the creation of the cosmological baryon asymmetry in the Two Higgs Doublet Model. We imagine a situation where the masses of the five Higgs particles and the two Higgs vevs are constrained by collider experiments, and demonstrate how the requirement of successful baryogenesis can be used to further constrain the remaining 4-dimensional parameter space of the model. We numerically compute the asymmetry within the scenario of Cold Electroweak Baryogenesis, which is particularly straightforward to simulate reliably.}


\begin{document}

\maketitle

\section{Introduction}
\label{sec:introduction}

With the discovery of what is expected to be the Higgs particle \cite{higgsmass}, the Standard Model of particle physics now provides a coherent and consistent theory of fundamental physics up to and including the electroweak scale. Although many phenomena in the realm of cosmology, such as inflation, dark matter and dark energy, are not addressed in this framework, all experimentally observed processes are very well described\footnote{We will consider the Standard Model to include right-handed neutrinos and a non-zero mass term for neutrinos. Hence the phenomenon of neutrino oscillations is considered part of the Standard Model.}. 

One central issue at the boundary between cosmology and particle physics is the origin of the baryon asymmetry observed in the Universe. A substantial effort has been made to link this phenomenon to electroweak scale physics \cite{EWBG1,EWBG2,EWBG3}, since it is the lowest energy where baryon number violation may occur. In combination with C-, P- and CP-violation and an out-of-equilibrium electroweak symmetry breaking transition, baryon asymmetry with the observed magnitude can indeed be produced. However, because Standard Model CP-violation is minute \cite{shaposhCP} (see however \cite{CPV1}) and the electroweak transition is a crossover in the Standard Model for the physical Higgs mass of 125-126 GeV \cite{rummukainen}, electroweak baryogenesis requires additional fields and interactions to exist.

Presumably the simplest way to achieve successful electroweak baryogenesis is to extend the scalar sector of the Standard Model by an additional field. This could be a $SU(2)$ singlet but probably the most popular extension is the Two-Higgs Doublet Model (2HDM) with an additional $SU(2)$ doublet. A number of different ``types'' exist, depending on how the two Higgs doublets couple to the fermions (see \cite{2hdmreview} for a recent review). The most general Higgs potential contains 14 (real) parameters, including up to two CP violating phases. In addition, because of the richness of the vacuum structure CP may be spontaneously broken. 

Restricting to a sub-class of models with only 10 real parameters, we ask the question whether the observed baryon asymmetry can be used to constrain the parameter space, complementing direct collider experiments. We expect that masses (4 different, of which we know the lightest) and vevs (2, of which we know one) are the easiest to measure, and so we will imagine that in future these are constrained, leaving a 4-dimensional less accessible subspace. Our aim here is to show how one may in principle sweep through this subspace and potentially use the observed baryon asymmetry to pin down the allowed parameter region.

The 2HDM doublet can accommodate a strong first order phase transition, but we will consider a different scenario, where electroweak symmetry breaking is a cold spinodal transition \cite{CEWBAG1,CEWBAG2,CEWBAG3,CEWBAG4}. This is a viable alternative to the standard ``Hot'' scenario \cite{EWBG2}, but for our purpose here, its main virtue is that it is straightforward to simulate numerically from first principles. 
Cold electroweak baryogenesis may be realized as a result of coupling to another scalar field, which may \cite{inflation} or may not \cite{notinflation,servant} be the inflaton. We demonstrated in \cite{paper01} through direct numerical simulations that a baryon asymmetry is indeed produced, as a result of the interplay between the explicit CP/C violation in the Higgs potential and the C- and P-violating (but CP-conserving) gauge-fermion interactions. It turns out that when simulating a bosonized version of the theory, it is necessary to include the P-breaking of fermions, and we did this through an effective higher order bosonic interaction term, parameterized by a coefficient $\delta_{\rm C/P}$. Although this coefficient can in principle be computed analytically, this is a very non-trivial task and we chose to keep it as a free parameter. We found that in order for the observed asymmetry to be reproduced, we need $\delta_{\rm C/P}\simeq (\textrm{2 to 3})\times 10^{-4}$ or larger.

The paper is structured as follows: In section \ref{sec:model} is a brief introduction of the bosonized 2HDM. In section \ref{sec:alpha} we parametrize the 4-dimensional parameter space in terms of a field transformation, two angles and one mass scale. The numerical results are presented in section \ref{sec:fullnum}, where we for a given range of parameters, and using lattice simulations in real-time, directly compute the baryon asymmetry in the bosonized electroweak sector. We conclude in section \ref{sec:conclusion}. In Appendix \ref{app:higgs}, we further discuss the parametrization of the neutral Higgs masses.

\section{The 2HDM}
\label{sec:model}

The 2HDM is defined through the continuum action
\ba
\label{eq:action}
S=-\int d^3x\, dt\,\bigg[
\frac{1}{4g^2}\textrm{Tr}\, F_{\mu\nu}F^{\mu\nu}+
(D_\mu\phi_1)^\dagger D^\mu\phi_1+(D_\mu\phi_2)^\dagger D^\mu\phi_2+V(\phi_1,\phi_2)
+\mathcal{L}_\textrm{C/P}
\bigg],\nonumber\\
\ea
where we use the metric $\eta={\rm diag}(-+++)$, $\phi_{1,2}$ are SU(2) doublets with hypercharge $+1$ and $F^{\mu\nu}$ is the field strength tensor of the gauge field. We will ignore the SU(3) and U(1) gauge fields. The covariant derivative is
$D_\mu\phi_i=(\partial_\mu+iA_\mu)\phi_i$ and the potential is in all generality \footnote{We here correct an error in \cite{paper01} in the normalization of the coefficients. The results obtained there were based on the conventions presented here.}
\ba\label{equ:V}
V(\phi_1,\phi_2)&=&-\frac{\mu_{11}^2}{2}\phi_1^\dagger\phi_1-\frac{\mu_{22}^2}{2}\phi_2^\dagger\phi_2-\frac{\mu^2_{12}}{2}\,\phi_1^\dagger\phi_2-\frac{\mu^{2,*}_{12}}{2}\phi_2^\dagger\phi_1\nonumber\\
&&+\frac{\lambda_1}{2}(\phi_1^\dagger\phi_1)^2+\frac{\lambda_2}{2}(\phi_2^\dagger\phi_2)^2+\lambda_3(\phi_1^\dagger\phi_1)(\phi_2^\dagger\phi_2)+\lambda_4(\phi_2^\dagger\phi_1)(\phi_1^\dagger\phi_2)\nonumber\\
&&+\frac{\lambda_5}{2}(\phi_1^\dagger\phi_2)^2+\frac{\lambda_5^*}{2}(\phi_2^\dagger\phi_1)^2
+\lambda_6(\phi_1^\dagger\phi_1)(\phi_1^{\dagger}\phi_2)
+\lambda_6^*(\phi_1^\dagger\phi_1)(\phi_2^{\dagger}\phi_1)\nonumber\\
&&+\lambda_7(\phi_2^\dagger\phi_2)(\phi_1^{\dagger}\phi_2)
+\lambda_7^*(\phi_2^\dagger\phi_2)(\phi_2^{\dagger}\phi_1).
\ea
The parameters $\lambda_{1,2,3,4}$ and $\mu^2_{11,22}$ are real and in general $\lambda_{5,6,7}$ and $\mu_{12}^2$ are complex. In this paper, we only study the 2HDM with a softly broken $Z_2$ symmetry, in which $\lambda_6=\lambda_7=0$ \cite{2hdmreview}. There is then only one independent CP violating phase. In the Standard Model as well as the 2HDM there is also CP-violation through the complex phase in the CKM mixing matrix. For the purpose of the present work, we will assume that the effective CP-breaking terms arising from this are negligible, although at very low temperatures, this may not be correct \cite{CPV1}.

We will take (\ref{eq:action}) to represent a bosonized version of the full theory, where fermions have been integrated out, and their effect is captured in a C- and P-breaking term given by \cite{turok}
\ba
\label{eq:CandP}
\mathcal{L}_{\rm C/P}=\frac{\delta_{\rm C/P}}{16\pi^2 m_W^2}i(\phi_1^\dagger \phi_2-\phi_2^\dagger \phi_1)\textrm{ Tr}\,F_{\mu\nu}\tilde{F}^{\mu\nu},
\ea
The Yukawa couplings and the mixing matrix is encoded in the real parameter $\delta_{\rm C/P}$, and it can in principle be computed from the model. The standard prescription in bosonized theories, which we will also adopt here, is then to infer the value of the baryon number $B$ through the anomaly equation
\ba
\label{eq:anomalyeq}
B(t)-B(0)=3[N_{\rm cs}(t)-N_{\rm cs}(0)],
\ea
where $N_{\rm cs}$ is the Chern-Simons number of the SU(2) gauge field.

The reason for including the term (\ref{eq:CandP}) is that, as demonstrated in \cite{paper01}, to generated a non-zero average Chern-Simons number, we need P-symmetry to be broken as well as CP-symmetry. It is easy to see that (\ref{eq:CandP}) conserves CP. 

 It turns out that in a finite temperature environment, the Higgs winding numbers $N_{\rm w}^{1,2}$ for the two Higgs fields, respectively, are much cleaner observables. At late times, the three agree, $N_{\rm w}^{1,2}=N_{\rm cs }$, and so we will identify the winding numbers at the end of the simulation to be the late time value for Chern-Simons number and hence the baryon asymmetry.

\section{Choices of the parameters}
\label{sec:alpha}

\subsection{The full parameter space of the 2HDM}
\label{sec:fullparam}

We will re-parametrize the 10-dimensional parameter space in the following way:
\begin{itemize}
\item {\bf Vacuum parameters (3):} $v$, $\beta$ and $\theta$.\\
Without loss of generality, we can parametrize the Higgs fields in terms of 2 complex and 4 real degrees or freedom as
\be
\phi_1=
e^{i\theta}\left(\begin{array}{c}
\phi_1^+\\
\left( v_1 + \eta_1 + i \chi_1\right)/\sqrt{2}
\end{array}\right),\qquad\phi_2=\left(\begin{array}{c}
\phi_2^+\\
\left( v_2 + \eta_2 + i \chi_2\right)/\sqrt{2}
\end{array}\right).
\ee
and
\bea
\chi_1 &=& \cos\beta G^0 - \sin\beta \eta^3,~~\chi_2 = \cos\beta \eta^3 + \sin\beta  G^0,\\
\phi_1^+ &=& \cos\beta G^+ - \sin\beta H^+,~~\phi_2^+ = \cos\beta H^+ + \sin\beta  G^+.
\eea
The vacuum is given by $G^{0,+}=\phi_{1,2}^+=\eta_{1,2,3}=0$, in terms of $v_1e^{i\theta}$ and $v_2$. We introduce $v$ and $\beta$ through
\ba
v_1 = v \cos\beta, \qquad v_2=v \sin\beta, \qquad v_2/v_1=\tan\beta.
\ea
Minimizing the Higgs potential gives three equations
\be
\f{\partial}{\partial v_1} V|_{v_1,v_2,\theta}=0,\quad \f{\partial}{\partial v_2} V|_{v_1,v_2,\theta}=0,\quad \f{\partial}{\partial \theta} V|_{v_1,v_2,\theta}=0,
\ee
with which we can replace three couplings/mass parameters by $\beta$, $\theta$ and $v$.

\item {\bf Higgs masses (4) :} $m_{1,2,3}$ and $m_\pm$.\\
There are five physical Higgs bosons: two form one charged field $H^\pm$ and the rest are mass eigenstates formed as linear combinations of the neutral fields $\eta_{1,2,3}$. We introduce the mass eigenvalues for these, $m_\pm$ and $m_{1,2,3}$, respectively, and these replace four other parameters (see also Appendix \ref{app:higgs}).
\item {\bf Neutral Higgs mixing angles (2) :} $\alpha_1$, $\alpha_2$.\\
As discussed in Appendix \ref{app:higgs}, the mass matrix of the neutral Higgs modes is in general not diagonal in the fields $\eta_{1,2,3}$, but it can be diagonalized through three mixing angles $\alpha_{1,2,3}$. Only two of these are independent, and we take $\alpha_3$ to be fixed through Eq.~(\ref{eq:alpha3}), which has 0, 1 or 2 solutions for a given set of $(\alpha_{1},\alpha_2)$.
\item {\bf A mass parameter (1) :}  $\mu^2=\textrm{Re}(\mu_{12}^2 e^{-i\theta})$.
\end{itemize} 
 
At the end of the day, the parameter set in the Higgs potential denoted by $\{\lambda\}$ is a function of the above 10 parameters. In the following discussion, for simplicity of notation we will use $\{\lambda\}[\ldots]$ with the ellipsis being some of the above parameters relevant for the discussions only.

\subsection{The subspace spanned by ($\alpha_{1}$, $\alpha_{2}$, $\theta$, $\mu$)}
\label{sec:4param}

As explained in the Introduction, we will assume that the 4 distinct masses and the two vevs have been determined (or at least constrained) by experiment, so that we can assign values to them:

\begin{itemize}
\item {\bf Vacuum parameters:}\\
The vev $v$ is known but not $\beta$ and we choose
\ba
v = 246~\text{GeV}, \qquad \tan\beta = 2.\label{equ:vac}
\ea
\item {\bf Higgs masses:}
The lowest neutral Higgs mass $m_1$ is fixed by experiment \cite{higgsmass}. Based on our choice of $\beta$ and experimental constraints \cite{2hdmreview}, we choose
\ba
m_1=125\,\textrm{GeV}, m_2=300\,\textrm{GeV}, m_3=350\,\textrm{GeV},m_{\pm}=400\,\textrm{GeV}.\label{equ:masses}
\ea
\end{itemize}

This leaves a 4-dimensional parameter space, spanned by $\alpha_{1}$, $\alpha_{2}$, $\theta$, and $\mu$.

\subsubsection{Symmetries}
\label{sec:symmetries}

Symmetries in the Higgs potential $V$ help us to further simplify our calculations. Since $V$ is real,
it follows that
\be
\{ \lambda \} \to \{ \lambda \}^*,\qquad \Phi_i  \xrightarrow{C} \Phi_i^*,\label{equ:complexConj}
\ee
is a symmetry. This imposes the relation between the sets of parameters $\{\lambda\}$,
\ba
\label{eq:alpha3symmetry}
\{\lambda\}[\alpha_1,\alpha_2,\alpha_3]=\left(\{\lambda\}[\alpha_1,-\alpha_2,\pi-\alpha_3]\right)^*,
\ea
which is equivalent to the charge conjugation of the bosonic fields according to (\ref{equ:complexConj}). Therefore, the generated baryon asymmetry flips sign when complex conjugating the parameter set $\{\lambda\}$, i.e.,
\bea
\label{eq:alpha3symmetry2}
&&N_{\rm cs}[\alpha_1,\alpha_2,\alpha_3]=-N_{\rm cs}[\alpha_1,-\alpha_2,\pi-\alpha_3],\nn\\
&&N_{\rm w}^{1,2}[\alpha_1,\alpha_2,\alpha_3]=-N_{\rm w}^{1,2}[\alpha_1,-\alpha_2,\pi-\alpha_3],
\eea
and so there is a redundancy between the upper and lower half-plane in $\alpha_1-\alpha_2$ space. 

Finally, the symmetry
\be
\phi_1 \to e^{-i \theta} \phi_1,\qquad\lambda_5\to e^{-2i\theta}\lambda_5,\qquad\mu_{12}^2 \to e^{-i\theta} \mu_{12}^2.
\label{sec:thetatrans}
\ee
will also be very useful. Using this transformation, one can easily see that
\be
\lambda_5[ \theta ] = e^{2i\theta} \lambda_5[ 0 ],\qquad\mu_{12}^2[ \theta ] =  e^{i\theta} \mu_{12}^2[ 0 ].\label{equ:theta}
\ee
Therefore, one can first find the parameter set  $\{ \lambda \}[0]$ and then obtain $\{ \lambda \}[\theta]$ by the above transformation. The physical Higgs masses are unchanged under such a transformation. Since $\mu$ is invariant under (\ref{equ:theta}), we need only consider varying the potential in the 3-dimensional parameter space spanned by $(\alpha_1, \alpha_2, \mu)$, and we get the $\theta$-direction for free.

The potential at different values of $\theta$ are equivalent, but with different field basis. However, the symmetry in (\ref{equ:theta}) is explicitly broken as soon as the scalar sector is coupled to fermions, or in our case the C-/P-violating term in (\ref{eq:CandP}) is included. Then different $\theta$ are physically distinct, as under the transformation (\ref{sec:thetatrans}),
\be
\phi_1^\dagger\phi_2-\phi_2^\dagger\phi_1\rightarrow e^{i\theta}\phi_1^\dagger\phi_2-e^{-i\theta}\phi_2^\dagger\phi_1.
\ee

\subsubsection{Basic constraints, maxima and saddle points}
\label{sec:maxsad}

For a given value of $\mu$, we now survey the whole $\alpha_1-\alpha_2$ plane. For each such pair, we accept/reject based on overall stability (potential is bounded from below), unitarity (tree-level Higgs-Higgs scattering amplitudes are smaller than unity), and whether the minimum found is a global minimum. Conditions for stability and unitarity are well-known and the interested reader is referred to \cite{2hdmreview} and references therein. For the requirement of the global minimum, we find all the other minima of the potential and establish that the chosen one is in fact the one with lowest potential energy. We also reject if there are no solutions for $\alpha_3$, and finally we reject if the potential has a minimum at $(\phi_1,\phi_2)=(0,0)$ (see below).

The origin $(\phi_1,\phi_2)=(0,0)$ is always a stationary point of $V$. For each of the surviving pairs $(\alpha_1,\alpha_2)$, we compute the eigenvalues of the mass matrix at the origin (not to be confused with $M^2$ of (\ref{eq:M2}), the neutral Higgs sector mass matrix in the minimum),
\ba
\mathcal{M}^2=-\frac{1}{2}\left(\begin{array}{cc}
\mu_{11}^2&\mu_{12}^2\\
\mu_{12}^{2,*}&\mu_{22}^2
\end{array}\right).
\ea
If both eigenvalues are negative, both Higgs fields will experience a spinodal transition, and we name this parameter point a {\it maximum}. If only one eigenvalue is negative (and the other positive), only one field goes spinodal, and we name the parameter point a {\it saddle point}\footnote{Note that since both fields acquire expectation values, eventually also the second field must undergo symmetry breaking, but then as a result of the first field going through its spinodal transition.}. If both eigenvalues are positive, no spinodal instability occurs and we reject the point. In principle, such a minimum could lead to tunneling and bubble nucleation on the way to symmetry breaking, but this returns us to standard electroweak baryogenesis, which we do not consider here (but see also \cite{servant}).

\begin{figure}
\begin{center}
\includegraphics[width=15cm]{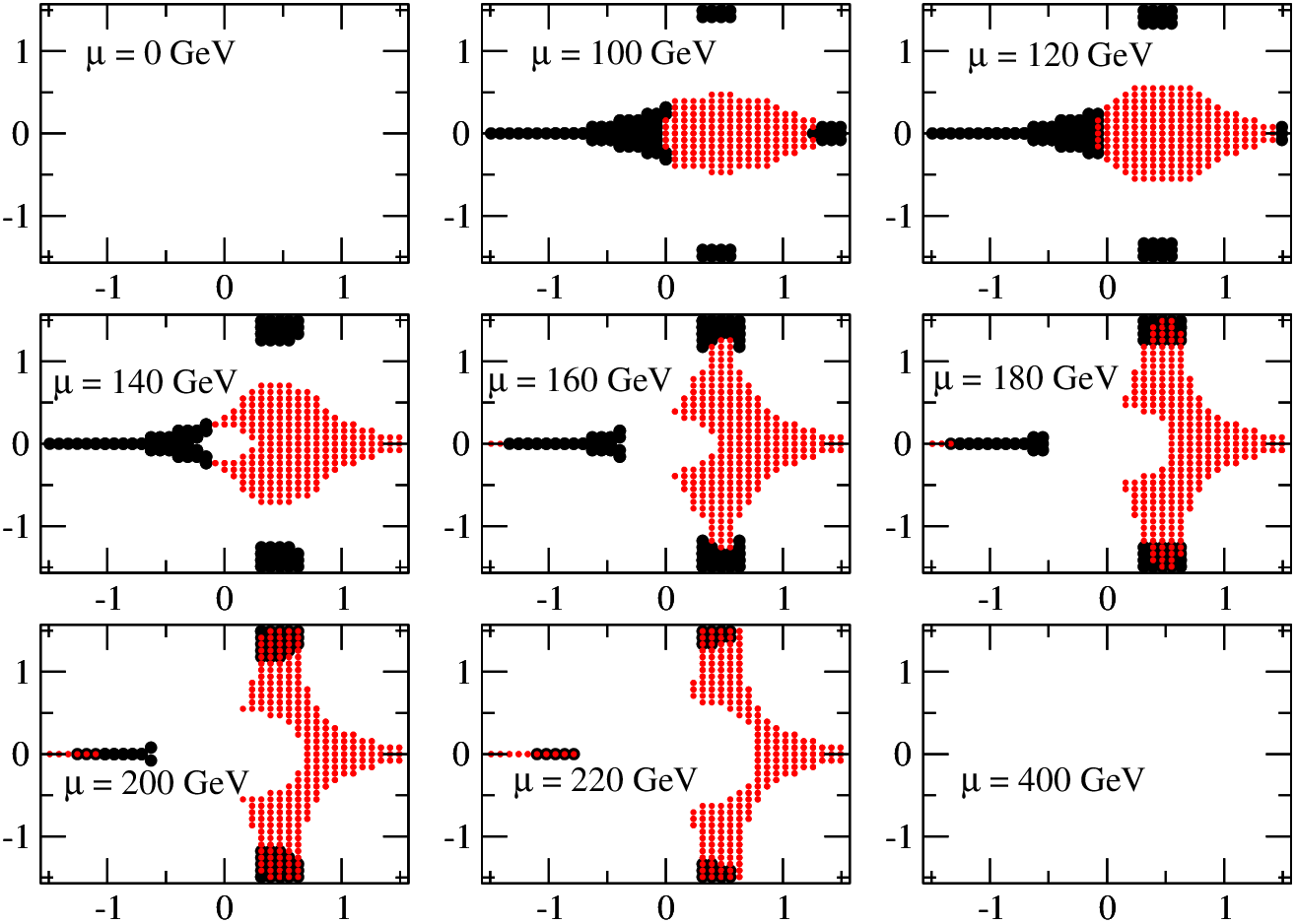}
\end{center}
\caption{The allowed values of $\alpha_{1,2}$ after all constraints have been applied, for different values of $\mu$. Black dots are maxima, red dots are saddle points. Two superposed points refer to the two different allowed values of $\alpha_3$. We perform simulations at every second allowed point at $\mu=100\,\text{GeV}$.}\label{fig:alphadivmu}
\end{figure}

Fig.~ \ref{fig:alphadivmu} shows the $\alpha_1-\alpha_2$ plane for various values of $\mu$, where we have sampled points with a spacing of $\pi/40$. We have indicated maxima by fat black dots, and saddle points by smaller red dots. The rest of the parameter space has been discarded for one of the reasons explained above. Where a red and a black dot are superposed, this corresponds to the two values of $\alpha_3$, and that these give a maximum and a saddle point, respectively. The lines $\alpha_2=0,\pm\pi/2$ have zero CP-violation, and can therefore not provide baryogenesis.

At zero $\mu$, no choice of $\alpha_{1,2}$ survives the constraints.  For small, but non-zero $\mu$, the allowed region is close to the $\alpha_1$-axis. For $\mu=100\,$GeV, about a third of the off-axis points are maxima, the rest are saddle points. As $\mu$ is further increased, the allowed region spreads out to a band near $\alpha_1=0.5$ which reconnects around the circle in the $\alpha_2$-direction. A ``hole'' also opens up around the origin. At the largest $\mu$, all off-axis points are saddle points, and by $\mu=400\,$ GeV, no points survive. Interestingly, by far the most important constraint is that the minimum should be the global minimum. All but a few of the discarded points fail in this respect. We expect that a similar picture arises for other choices of $m_{1,2,3,\pm}$ with the allowed region shifted accordingly in $\mu-\alpha_{1}-\alpha_{2}$-space.

We now turn to our numerical lattice simulations, where we have computed the baryon asymmetry for all the allowed parameter space for $\mu=100\,$GeV, top middle of Fig.~\ref{fig:alphadivmu}, but with a coarser spacing of $\pi/20$. 

\section{Numerical results: }
\label{sec:fullnum}

\begin{figure}
\begin{center}
\includegraphics[width=7cm]{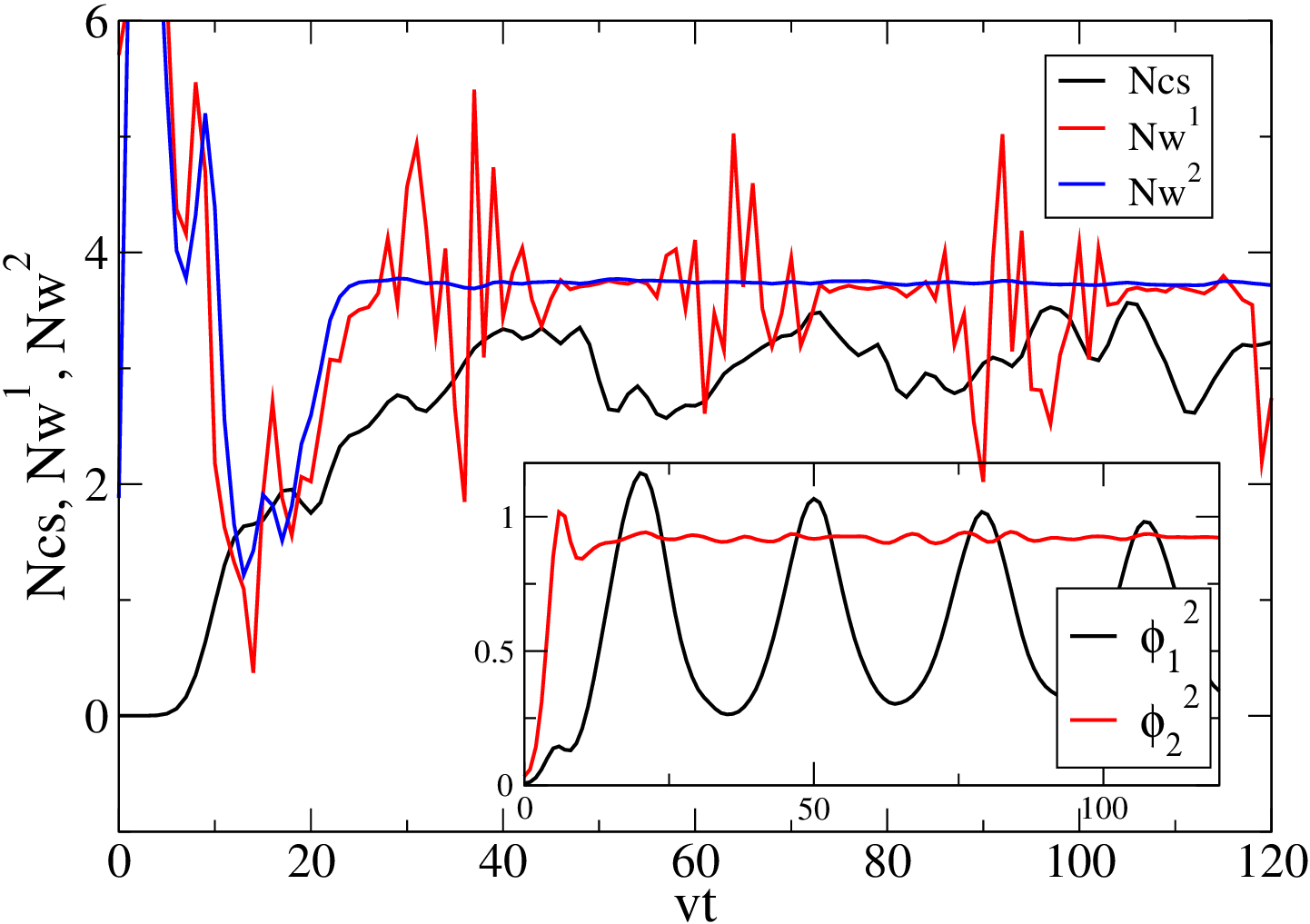}
\includegraphics[width=7cm]{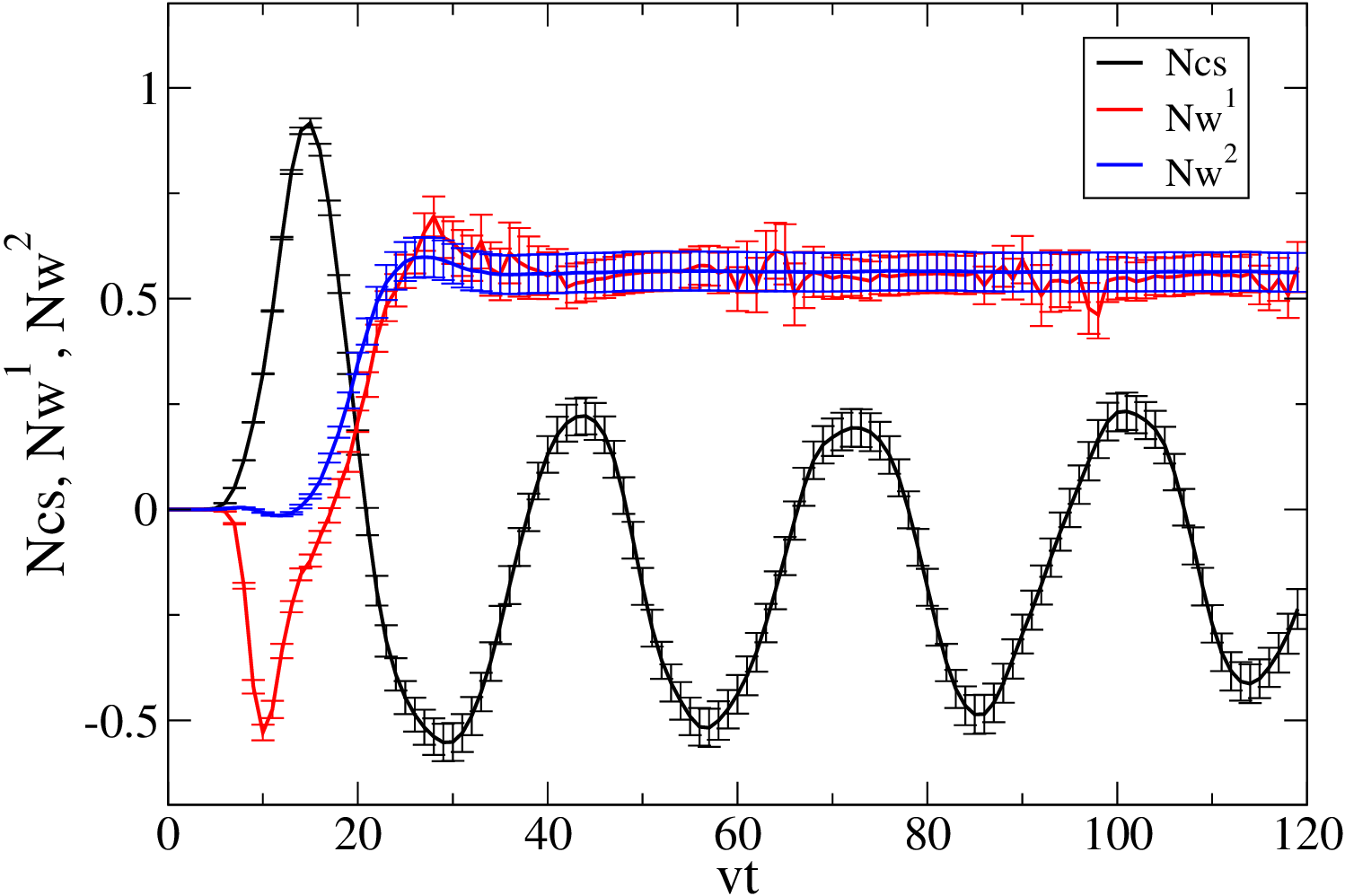}
\end{center}
\caption{Left: The Chern-Simons number and Higgs winding numbers in a single initial field realization. The inset shows the Higgs field expectation values squared. Right: The Chern-Simons number and Higgs winding numbers averaged over the ensemble.}\label{fig:averageexample}
\end{figure}

The action (\ref{eq:action}) is discretized on a lattice and the classical equations of motion derived and solved numerically. Starting from a zero-temperature initial condition, we study the evolution of the system through the spinodal transition. Observables are averaged over a statistical ensemble of initial realizations, which is by hand C-, P- and CP-symmetric (for details, see \cite{paper01}). The baryon asymmetry is inferred from the anomaly equation (\ref{eq:anomalyeq}). 

In fact, because we are initially very far from equilibrium,  Chern-Simons number is not a very clean observable, since it is in general non-integer and can exhibit large oscillations at intermediate times. Instead, we consider the Higgs winding number, which coincides with Chern-Simons number at late times, is integer throughout and settles much earlier into its late-time value. Since we have two Higgs fields, we also have two winding numbers, both of which will eventually match Chern-Simons number, and we identify the late-time value by the time at which the two agree, irrespective of the value of the Chern-Simons number.

In \cite{paper01}, we studied the dependence of the baryon asymmetry on the strength of C-/P-violation. In the present paper, we fix $\delta_{C/P}=-21$ which is close enough to the linear regime that we can interpolate to smaller values \cite{paper01} ($\delta_{\rm C/P}=0$ gives zero asymmetry by construction). In this way, we can investigate the significance of CP violation by studying the dependence of the baryon asymmetry on $\alpha_{1,2,3}$ and $\theta$. 

Fig.~\ref{fig:averageexample} (left) shows the evolution of winding numbers and Chern-Simons number for one particular configuration for a particular choice of $(\alpha_1,\alpha_2,\alpha_3)=(0,\pi/10,1.138)$. Winding number has clearly settled, while Chern-Simons number is still catching up. In the inset, we show the Higgs field expectation values squared, of which one settles very rapidly, and one keeps oscillating for a long time, and with large amplitude. This is because the potential around the minimum is steep in one direction and shallow in the other.

In Fig.~\ref{fig:averageexample} (right) we show the average winding number and Chern-Simons number, averaged over 100 sets of 4 conjugate configurations. Most of the high-frequency noise in the winding numbers has been averaged out, and the two nicely settle at a common value, quite early on in the evolution. By $vt=30$, symmetry breaking is complete. We also see that statistical errors are well under control at this size of ensemble. 

The average Chern-Simons number, however, has certainly not settled to its equilibrium value. Two effects are at work here: There is a net shift downwards, which is a transient non-equilibrium effect. We checked, by running for three times as long, that eventually the Chern-Simons number settles to the winding number value.

The second effect is a large-amplitude oscillation, and is the result of the C-/P-violating driving force being given by the oscillating Higgs field vevs. We see from the figure that the oscillation has the same frequency as the Higgs field oscillations, but are shifted by a phase. This follows from considering the C-/P-violating term as 
\ba
S_{\rm C/P}\propto \delta_{\rm C/P}\,\textrm{Im}[\phi_1^\dagger\phi_2]\, \partial_t N_{\rm cs},
\ea
which holds approximately for almost homogeneous Higgs fields. By partial integration, this term can be considered a time dependent driving force or chemical potential for Chern-Simons number, with magnitude $\propto \delta_{\rm C/P}\partial_t(\textrm{Im}[\phi_1^\dagger\phi_2])$. The reason why this second effect is not washed out by the averaging procedure is that all members of the ensemble experience (roughly) the same oscillation frequency and phase of the driving force, since the Higgs oscillation is almost universal, configuration by configuration. Therefore, although other configuration-specific effects average out to give a small baryon asymmetry, the driven oscillation is common to all configuration and survives the averaging process. At late times, the Higgs fields will also stop oscillating, and the driving force will disappear. But even at these early times, the coherent oscillation has no impact on the average winding numbers, which we therefore take as our measurement of the generated baryon asymmetry.

\subsection{Symmetry under $C$ and $\lambda\rightarrow \lambda^*$}
\label{sec:symC}

\begin{figure}
\begin{center}
\includegraphics[width=7cm]{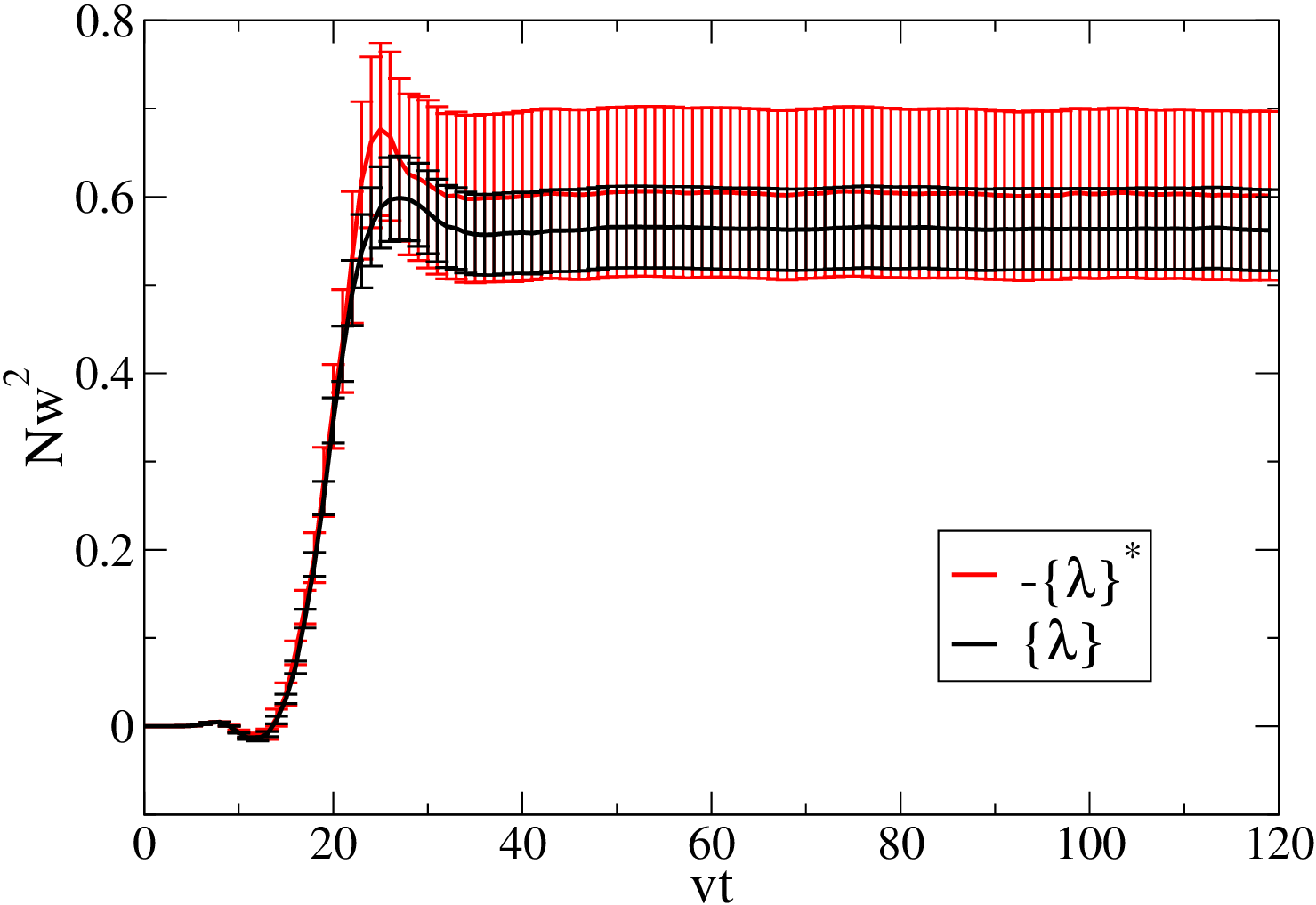}
\end{center}
\caption{Higgs winding number for a parameter set $\{ \lambda \}$ and for $\{ \lambda \}^*$, with an overall flipped sign. Here, $(\alpha_1, \alpha_2, \mu, \theta)=(0, \f{\pi}{10}, 100\,\text{GeV}, -1.27)$ and the results are averaged over an ensemble of $4\times25$ configurations. }\label{fig:Csym}
\end{figure}
In Fig.~\ref{fig:Csym}, we demonstrate explicitly that the symmetry (\ref{eq:alpha3symmetry2}) holds, by simply computing the asymmetry for a parameter set $\{\lambda\}$ and its complex conjugate, and then flipping the sign of the resulting asymmetry. We see that the agreement is very good (within statistical error bars). Hence we find the advertised redundancy between positive and negative values of $\alpha_2$.

\subsection{Dependence on $\theta$}
\label{sec:theta}

\begin{figure}
\begin{center}
\includegraphics[width=7cm]{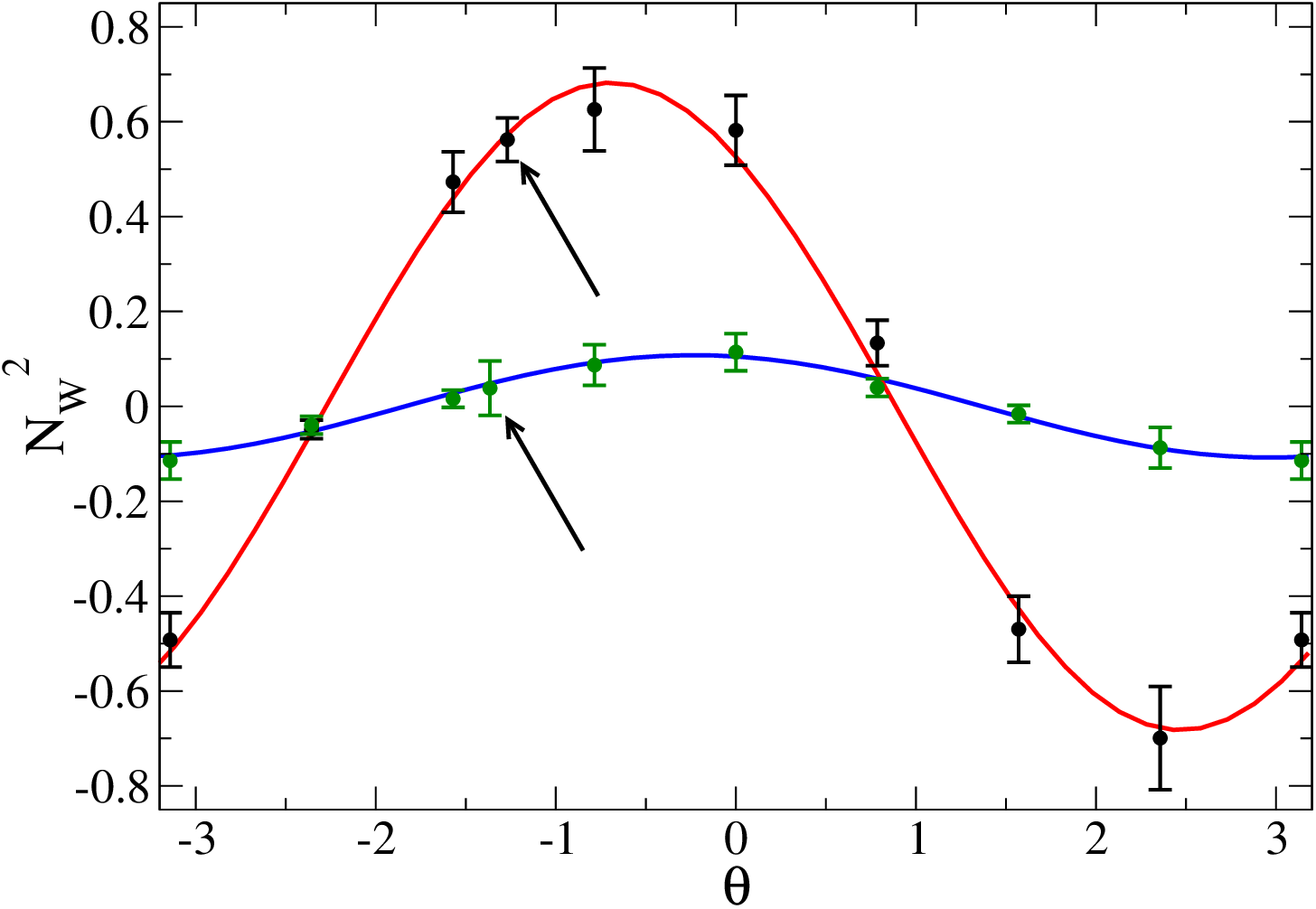}
\end{center}
\caption{The dependence of the winding number asymmetry on $\theta$ for $(\alpha_1, \alpha_2, \mu)=(0, \f{\pi}{10}, 100\,\text{GeV})$ (black and red) and $(\f{2\pi}{5}, \f{\pi}{20}, 100\,\text{GeV})$ (green and blue).}\label{fig:theta}
\end{figure}

The transformation (\ref{sec:thetatrans}) allows us, from a given set of parameters $\{\lambda\}$, to generate a whole set of identical potentials, but where the minimum is rotated to $v_1 e^{-i\theta}$ for any value of $\theta$. We generate $\{\lambda\}$ at $\theta=0$ using the constraint $\textrm{Im}\,\mu_{12}^2=v_1v_2\textrm{Im}\lambda_5$, and from the point of view of CP-violation, all values of $\theta$ are equivalent. But once we couple to C-/P-violation, the potentials are distinct. In the vacuum, we have
\be
\mathcal{L}_{\rm C/P}\propto \delta_{\rm C/P}2\,v_1\,v_2\sin(\theta)\textrm{ Tr}\,F_{\mu\nu}\tilde{F}^{\mu\nu},
\ee
and so were we in vacuum throughout the transition, there would be no asymmetry generated at $\theta=0$. And naively, one would expect the asymmetry to be proportional to $\sin\theta$. 

Fig.~\ref{fig:theta} shows the asymmetry in $N_{\rm w}^2$ as a function of $\theta$ for the parameter sets $\{\lambda\}(\alpha_1, \alpha_2, \mu)=(0, \f{\pi}{10}, 100~~\text{GeV})$ (black dots) and  $(\f{2\pi}{5}, \f{\pi}{20}, 100\,\text{GeV})$ (green dots). The arrows indicate the values of $\theta$ corresponding to real $\lambda_5$, the criterion we will use for most of our simulations below. We have fit with a form $A \sin(\theta+\delta\theta)$, and find beautiful agreement, but with a non-zero $\delta\theta=2.26$ (red line) and $\delta\theta=1.8$ (blue line). As a result, even at $\theta=0$, an asymmetry is generated during the transition where $\theta$ is different from its vacuum value.  

This is a result of the Higgs fields rolling down the potential in a spinodal transition, where both the length and the phase of the fields vary locally, until they finally settle near their vacuum values. The surprising result is perhaps that the simple $\sin\theta$ form is preserved, and that the out-of-equilibrium stage is encoded in the $\theta$-dependence as an overall shift of the phase, $\delta\theta$.

But this also means that the asymmetry vanishes at $\theta=-\delta\theta$ (and $\theta=\pi-\delta\theta$), and that the overall sign of the asymmetry varies in this simple way with $\theta$, presumable for any set $\{\lambda\}$. We do not know of an obvious way of determining $\delta\theta$ apart from through the simulations. On the other hand, since we can parameterize the dependence through $A$ and $\delta\theta$, we only need simulations at two  values of $\theta$, say $\theta=0$ and $\theta=\pi/2$. Then we simply have
\be
\tan(\delta\theta)=\frac{N_{\rm w}(\theta=0)}{N_{\rm w}(\theta=\pi/2)},\qquad A=\frac{N_{\rm w}(\theta=0)}{\sin(\delta\theta)},
\ee
from which one can find $\delta\theta$ and then $A$. 

In the following, we have for each $\{\lambda\}[0]$ rotated to the value of $\theta$ that gives a real $\lambda_5$. This choice is arbitrary, and can as we have seen with a comparable amount of additional some numerical effort be extended to a complete $\theta$-dependence. 

\subsection{Dependence on $\alpha_{1,2}$}
\label{sec:alphadep}

\begin{figure}
\begin{center}
\includegraphics[width=7cm]{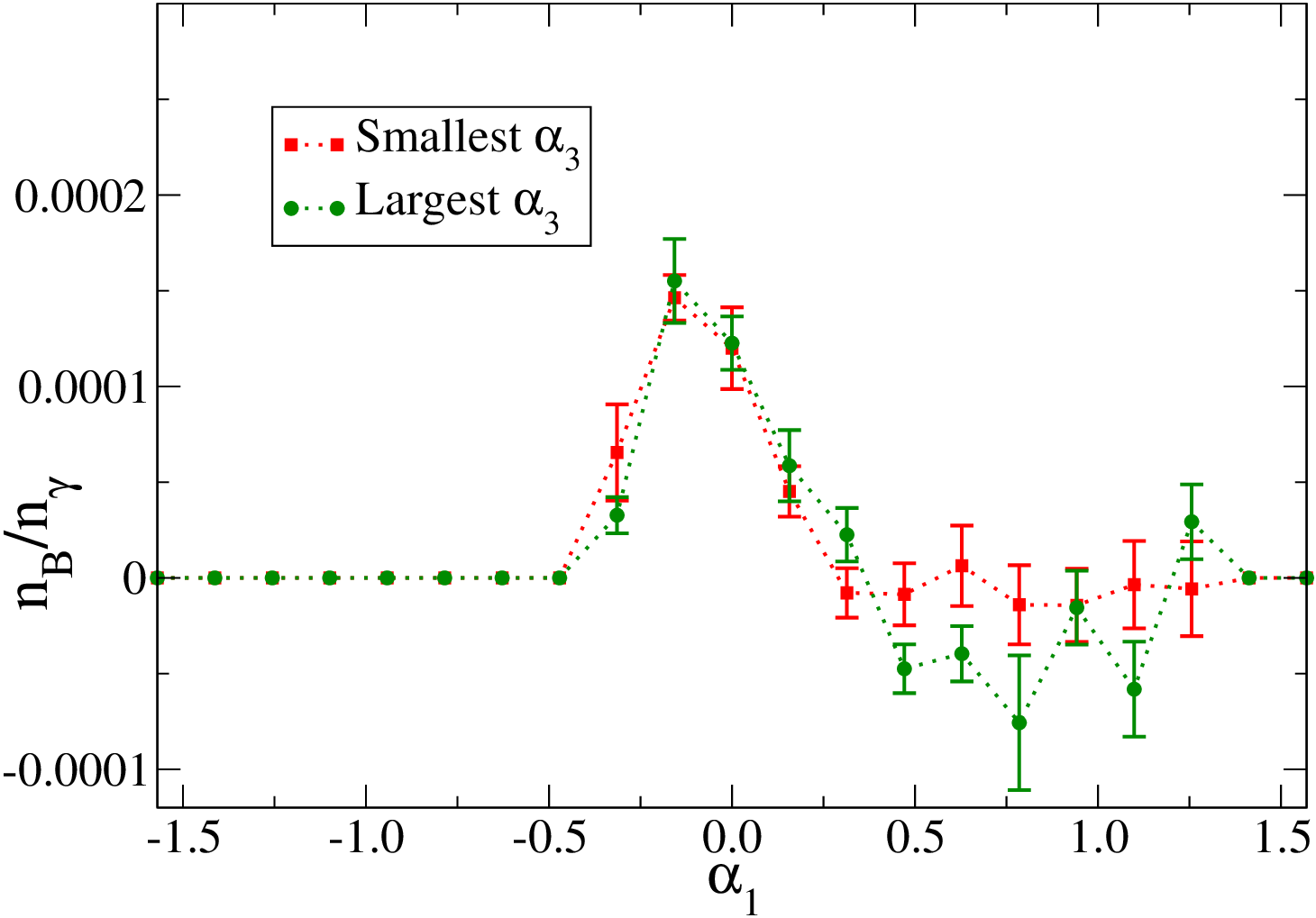}
\includegraphics[width=7cm]{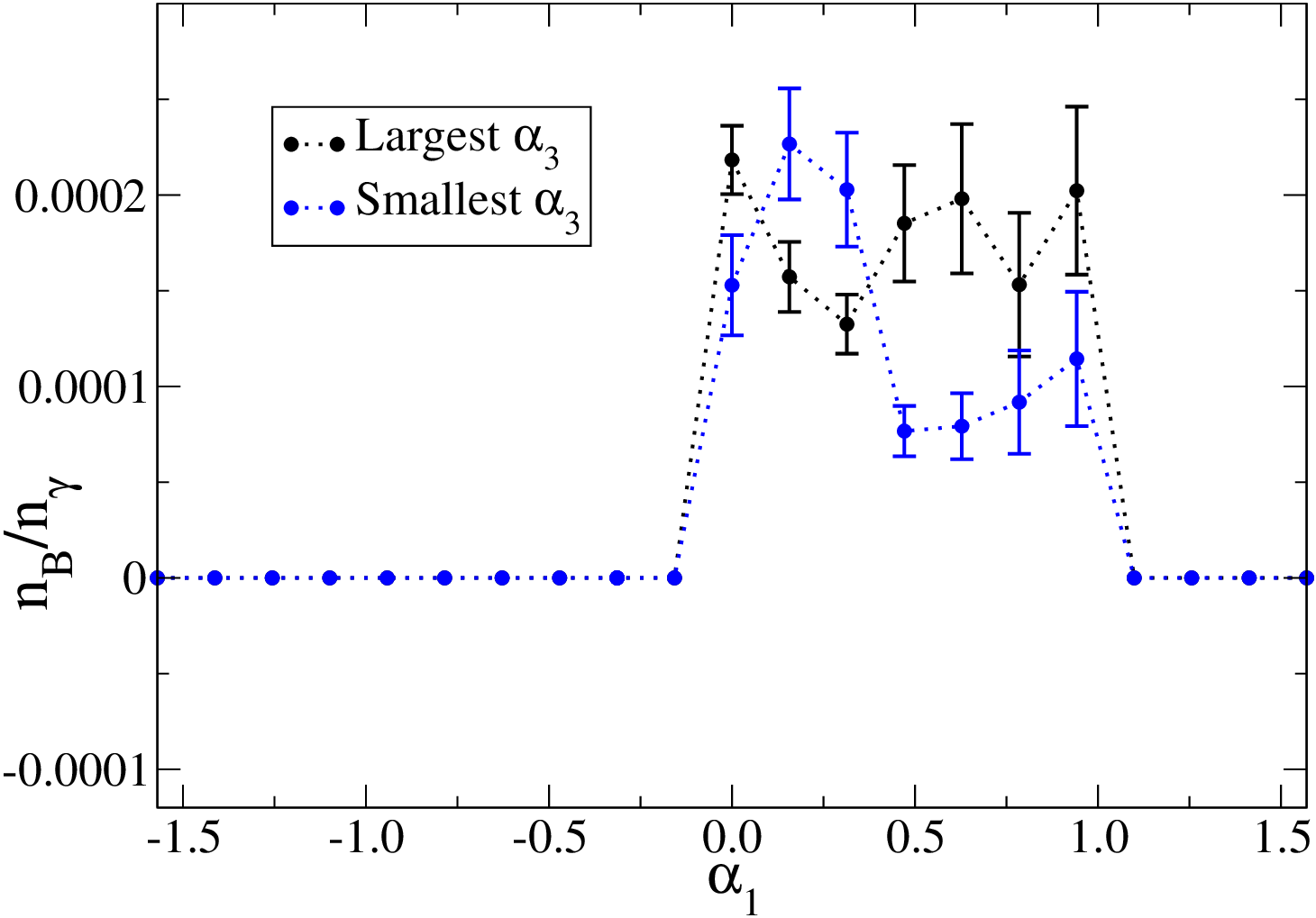}
\end{center}
\caption{The final baryon asymmetry as a function of $\alpha_1$ for $\alpha_2=\pi/20$ (left) and $\pi/10$ (right).}\label{fig:alpha12}
\end{figure}

Fig.~\ref{fig:alpha12} is the baryon asymmetry as a function of $\alpha_{1}$ at $\mu=100\,$GeV for $\alpha_2=\pi/20$ (left) and $\pi/10$ (right). We have used the conversion from winding number to baryon number
\ba
\frac{n_B}{n_\gamma}=1.2\times 10^{-4}\left(\frac{V(0,0)-V(v_1,v_2)}{v^4}\right)^{-3/4}\times \langle N_{\rm w}^{1,2}\rangle.
\ea 
This assumes that the total potential energy is distributed onto all the Standard Model degrees of freedom with masses less than $m_{\rm w}$, giving the reheating temperature and the photon number density $n_\gamma$.

As discussed in the previous section, the $\alpha_2>0$ and $\alpha_2<0$ regions are related by Eq.~(\ref{eq:alpha3symmetry})
Since there are at most two allowed values of $\alpha_3$ at each grid point in the $(\alpha_1, \alpha_2)$ plane, we finally need to perform our numerical simulation using 39 sets of parameters. 

At $\alpha_2=\pi/20$ (left) we see a clear peak close to $\alpha_1=0$, which gradually decreases toward the edges of the allowed parameter region. For large $\alpha_1$ the results are roughly consistent with zero. The two values of $\alpha_3$ happen to give very similar results within errors. Note that connecting the largest/smallest $\alpha_3$-results by curves is an arbitrary choice to guide the eye. At $\alpha_2=\pi/10$ (right) we first of all observe that the two values of $\alpha_3$ do not agree as well, although they are still within a factor of two. The asymmetry is larger than for $\alpha_2=\pi/20$, and shows no sign of smoothly going to zero at the edge of the parameter range. This may be a result of the coarse resolution in $\alpha_1$. Again, the connecting curves are just to guide the eye. We see that the maximum value is again attained near $\alpha_1=0$, although now a peak structure is less clear. We should also keep in mind the $\theta$-dependence, and that the results shown here could be at any point in the period of the $\sin\theta$ behaviour. 

The small remaining parameter range at large $\alpha_2$ (see Fig.~\ref{fig:alphadivmu}) gives baryon asymmetries of roughly the same size, and are by no means suppressed compared to small $\alpha_2$. We should also mention that we checked that the magnitude of the asymmetry is not in a simple way correlated with the determinant or eigenvalues of the mass matrix $\mathcal{M}^2$ at the origin, and in particular whether we start at a maximum or a saddle point. There is also no simple correlation with the phase of $\mu_{12}^2$ or of $v_1$. We did, however find a weak correlation in the combined $\textrm{Im}(v_1)$-$(V(0,0)-V(v_1,v_2))$-plane, suggesting that a deep potential drop and large CP-violation gives a large baryon asymmetry. This is perhaps not unexpected, but is surprisingly difficult to confirm. Clearly, the complicated non-linear dynamics does not allow for such simple conclusions about the generated asymmetry. 

\section{Conclusion and outlook}
\label{sec:conclusion}

We have outlined a practical parametrization of the 4-dimensional parameter space in the 2HDM resulting from fixing masses and vevs. We have seen that imposing a number of general consistency criteria, a finite region in $\mu-\alpha_1-\alpha_2$ survives, and this can be extended to the $\theta$-direction by a simple phase change transformation. 

The parameter space is a hyper-cylinder with $\mu>0$ and three angles $\alpha_1$, $\alpha_2$ and $\theta$, with the additional redundancy that $\alpha_2>0$ and $\alpha_2<0$ are connected. On the other hand each set of $\alpha_{1,2}$ has up to two solutions for $\alpha_3$. When the bosonic sector is coupled to fermions, different values of $\theta$ are physically distinct, but the asymmetry seems to follow a form $A\sin(\theta+\delta\theta)$. $\delta\theta$ is a priori unknown, but can be found numerically by using two different values of $\theta$, say spaced by $\pi/2$. Finally, it seems that the extent in $\mu$ is finite and determined by the overall scale of the fixed Higgs masses.

The maximal asymmetry we found is at the point $(\alpha_1,\alpha_2,\alpha_3)=(\pi/20,\pi/10,1.333)$, and is
\ba
\left(\frac{n_B}{n_\gamma}\right)_{\rm max}=-1.1\times 10^{-5}\times \delta_{\rm C/P},
\ea
so that in order to reproduce that observed asymmetry of $\sim 6\times 10^{-10}$, we require
\ba
\delta_{\rm C/P}\simeq (5\textrm{ to }6) \times 10^{-5}.
\ea
We note that the maximal asymmetry is a factor 3 or 4 larger than what we found in \cite{paper01} at one particular parameter point, which therefore was not a particularly unique case.  Also, there are parts of the allowed parameter space that give vanishing asymmetry. Computing $\delta_{\rm C/P}$ from first principles could therefore potentially rule out regions of 2HDM parameter space, under the assumption that baryogenesis originates from a cold spinodal transition involving two Higgs field.

Although a similar programme could be attempted for other scenarios of baryogenesis, these are less amenable to a direct, quantitative computation. Leptogenesis is a multi-stage process, generation of lepton asymmetry, thermalization, sphaleron processes, freeze-out. And ``Hot'' electroweak baryogenesis involves the nucleation of bubbles, their interaction with the plasma and again sphaleron processes. Cold electroweak baryogenesis offers a practicable testing groud for this kind of parameter scans.

Given the numerical effort involved in the present work (of order $10^5$ CPU hours on a standard linux cluster), it is difficult to scan through the currently allowed parameter space, including the remaining Higgs masses and $\tan \beta$. But a complete sweep of the 4-dimensional parameter space can be done with about a factor of 10-100 more computing power, which is easily within reach of current supercomputers. And hopefully, the coming years of LHC-experiments at the electroweak energy scale will constrain the viable range of masses and vevs, or even discover additional scalar particles. When this happens, it would be natural to revisit the scenario considered here, and use the baryon asymmetry to narrow down the range of experimentally less accessible parameters.

\acknowledgments
We would like to thank Aleksi Vuorinen, Tomas Brauner and Olli Taanila for helpful discussions. B.W. was supported by the Humboldt foundation through its Sofja
Kovalevskaja program. A.T. was supported by the Carlsberg Foundation and the Villum Kann Rasmussen Foundation.

\appendix

\section{Masses of the physical Higgs bosons}
\label{app:higgs}

The physical Higgs bosons are defined by the mass eigenstates, which can be found by writing 
\be
\phi_1=
e^{i\theta}\left(\begin{array}{c}
\phi_1^+\\
\left( v_1 + \eta_1 + i \chi_1\right)/\sqrt{2}
\end{array}\right),\qquad \phi_2=\left(\begin{array}{c}
\phi_2^+\\
\left( v_2 + \eta_2 + i \chi_2\right)/\sqrt{2}
\end{array}\right).
\ee
There are in total eight (real) fields but three of them contribute to the degrees of freedom of the massive gauge bosons $W^\pm$ and $Z^0$ after spontaneous symmetry breaking. Let us write
\bea
&&\chi_1 = \cos\beta G^0 - \sin\beta \eta^3,~~\chi_2 = \cos\beta \eta^3 + \sin\beta  G^0,\label{equ:chi}\\
&&\phi_1^+ = \cos\beta G^+ - \sin\beta H^+,~~\phi_2^+ = \cos\beta H^+ + \sin\beta  G^+,\label{equ:phi}
\eea
where $G^0$ and $G^\pm$ ($G^-$ is the complex conjugate of $G^+$) are Goldstone bosons to be swallowed up by the gauge bosons. Inserting (\ref{equ:chi}) and (\ref{equ:phi}) into (\ref{eq:action}), we have for the remaining degrees of freedom
\be
V = \f{1}{2} \eta M^2 \eta^T + m_{\pm}^2  H^- H^+ + \text{interaction terms},
\ee
where (using $s_\beta\equiv \sin\beta$, $c_\beta\equiv\cos\beta$, $t_\beta\equiv\tan\beta$)
\be
m_{\pm}^2 = \frac{1}{2} \left[-v^2 (\lambda_4 + \hat{\lambda}_5^\textrm{Re} )+ {\hat{\mu}_{12}^{2,\textrm{Re}}} /(c_\beta s_\beta) \right],\label{equ:mHp}
\ee
with
\be
\hat{\lambda}_5\equiv \lambda_5 e^{-2 i \theta} \equiv \hat{\lambda}_5^\textrm{Re}+i \hat{\lambda}_5^\textrm{Im} ,~~\text{and}~~\hat{\mu}_{12}^2 \equiv \mu_{12}^2 e^{-i\theta}\equiv {\hat{\mu}_{12}^{2,\textrm{Re}}} + i {\hat{\mu}_{12}^{2,\textrm{Im}}}. 
\ee
$H^{\pm}$ are the charged Higgs bosons with mass $m_{\pm}$. The three neutral physical Higgs bosons are obtained by diagonalizing $M^2$, which reads explicitly
\ba
\label{eq:M2}
M^2=\left(
\begin{array}{ccc}
v^2\lambda_1(c_\beta)^2+\frac{1}{2}\hat{\mu}_{12}^{2,\textrm{Re}}t_\beta&
\frac{1}{2}\hat{\mu}_{12}^{2,\textrm{Re}}+v^2\lambda_{345}c_\beta s_\beta&
-\frac{1}{2}\hat{\mu}_{12}^{2,\textrm{Im}}/c_\beta\\
\frac{1}{2}\hat{\mu}_{12}^{2,\textrm{Re}}+v^2\lambda_{345}c_\beta s_\beta&
\frac{1}{2}\hat{\mu}_{12}^{2,\textrm{Re}}/t_\beta+v^2\lambda_2(s_\beta)^2&
-\frac{1}{2}\hat{\mu}_{12}^{2,\textrm{Im}}/s_\beta\\
-\frac{1}{2}\hat{\mu}_{12}^{2,\textrm{Im}}/c_\beta&
-\frac{1}{2}\hat{\mu}_{12}^{2,\textrm{Im}}/s_\beta&
\hat{\mu}_{12}^{2,\textrm{Re}}/s_{2\beta}-v^2\hat{\lambda}_5^\textrm{Re}
\end{array}
\right),
\ea
with $\lambda_{345}\equiv(\lambda_3+\lambda_4+\hat{\lambda}_5^\textrm{Re})$. Following Ref. \cite{osland1,osland2}, we introduce the rotational matrix
\be
R_{n}=\left(
\begin{array}{ccc}
 c_1 c_2 & c_2 s_1 & s_2 \\
 -c_3 s_1-c_1 s_2 s_3 & c_1 c_3-s_1 s_2 s_3 & c_2 s_3 \\
 -c_1 c_3 s_2+s_1 s_3 & -c_3 s_1 s_2-c_1 s_3 & c_2 c_3
\end{array}
\right),
\ee
such that,
\be
M^2 = R_n^T \text{diag}\left\{m_1^2, m_2^2, m_3^2\right\}R_n,\label{equ:M2n}
\ee
where $s_i\equiv \sin\alpha_i$ and $c_i\equiv \cos\alpha_i$ with $i = 1, 2, 3$, and the rotational angles $\f{\pi}{2}>\alpha_1, \alpha_2 \geq -\f{\pi}{2}$ and $\pi>\alpha_3 \geq 0$. There are 6 independent equations in (\ref{equ:M2n}), and one of them gives a constraint between the three rotational angles, which reads
\ba
\label{eq:alpha3}
({m_3/m_2})^2= \frac{ \sin(2 {\alpha_3}) \tan({\alpha_1}+\beta )-2 \sin({\alpha_2}) \left(({m_1/m_2})^2-\sin({\alpha_3})^2\right)}
{ \sin(2 {\alpha_3}) \tan({\alpha_1}+\beta )-2 \sin({\alpha_2})(1-\sin(\alpha_3)^2)},
\ea
and to which there are 0, 1 or 2 solutions for $\alpha_3$ for each pair $(\alpha_1,\alpha_2)$.



\end{document}